\documentclass{article}
\usepackage{amsmath}

\usepackage{arxiv}

\usepackage[utf8]{inputenc} 
\usepackage[T1]{fontenc}    
\usepackage{hyperref}       
\usepackage{url}            
\usepackage{booktabs}       
\usepackage{amsfonts}       
\usepackage{nicefrac}       
\usepackage{microtype}      
\usepackage{lipsum}		
\usepackage{graphicx}

\title{ Physical viability of traversable Finslerian wormholes with traceless fluid under conformal symmetry}

\author{Manjunath Malligawad$^1$ S. K. Narasimhamurthy$^{1*}$ Z. Nekouee$^2$ Rajesh Kumar$^3$\\$^1$Department of PG Studies and Research in Mathematics,
	Kuvempu University, Jnana Sahyadri, \\Shankaraghatta - 577 451, Shivamogga, Karnataka, India.\\$^2$School of Physics, Damghan University, Damghan, 3671641167, Iran.\\$^3$Department of Mathematics and Statistics,  DDU Gorakhpur University,\\ Gorakhpur, Uttaar Pradesh, India.\\
	\texttt{Corresponding author$^*$: nmurthysk@gmail.com} \\
}

\hypersetup{pdfkeywords={Traversable wormhole, Finsler gravity, Conformal symmetry,  Energy conditions, traceless fluid.}
}

\begin{document}
\maketitle

\begin{abstract}
The current study explores the novel potential of traversable wormhole solutions within the framework of Finsler geometry, incorporating conformal symmetry alongside traceless fluid dynamics. Using the Conformal Killing vector approach, we have discussed the wormholes based on traceless fluid within the intriguing framework of Finsler geometry. The field equations and the associated conformal factor are obtained specifically under the condition of conformal motion in Finsler geometry. Furthermore, we have successfully derived and examined the shape function, considering a range of values for the Finslerian parameter $\lambda$. Our investigation extends to fundamental physical characteristics such as proper radial distance, active mass function, and total gravitational energy, aiming to understand their influence on the traversability of the wormhole. The observation of energy condition violations provides evidence for the exotic matter's presence near the throat, reinforcing the assertion of the Finslerian wormhole's traversability.
\end{abstract}


\section{Introduction}\label{sec.1}

Wormholes (WHs) are fascinating solutions arising from General Relativity Theory and offer potential shortcuts for interstellar travel and could serve as time machines.
Initially conceived as pedagogical tools for teaching General Relativity, the idea of WHs gained momentum with the emergence of proposals for their actual existence. Imagine the universe as a vast fabric and WHs as secret tunnels threading through it, connecting far-off places.
These tunnels could be shortcuts, like secret passages in a giant maze, allowing us to hop from one point to another without traveling huge distances. Scientists are exploring these WHs, thinking about how they might stay open.
They are also looking at special fluids that behave uniquely in space. One way scientists are studying this is by using a mathematical idea. In scientific discussions, WHs are divided into static and dynamic categories. Static WHs maintain a constant radius at their throats, while dynamic WHs exhibit variability in the size or radius of this crucial area. Einstein and Rosen \cite{Einstein} were the first to propose a solution involving WHs by utilizing the concept of an event horizon termed Lorentzian WHs. Several years later, Morris and Thorne \cite{Morris} introduced the traversable WHs concept. Weyl, in \cite{Weyl}, introduced a wormhole model and analyzed mass implications by deriving Einstein's field equations. The challenge posed by exotic matter in WHs may be mitigated, by exploring the structure of developed gravity theories. These theories, commonly employed to address contemporary cosmological issues like dark energy and dark matter, provide a potential avenue to bypass the exotic matter requirement \cite{Saaidi, Singh}.
\par

Finsler geometry, a crucial geometric framework, has been considered a viable approach for studying WHs in this context. Finsler geometry, as an alternative to general relativity, treats the four-velocity vector as an independent variable, encompassing Riemann geometry as a particular case \cite{Bao}. Cartan \cite{Cartan} initiated the self-consistent Finsler geometry model in 1935, and the Einstein-Finsler equations for the Cartan d-connection were introduced in 1950 \cite{Horvath}. Over the years, various Finsler geometry models have been studied in physics applications \cite{Vacaru, Vacaru1, Schreck}. In the mid-1990s, Vacaru \cite{Vacaru2, Vacaru3} constructed relativistic models of Finsler gravity, deriving it from the low-energy limits of superstring/supergravity theories with N-connection structure. Vacaru et al. \cite{Vacaru4, Stavrinos, Rajpoot} studied the anholonomic frame deformation method (AFDM) using Finsler geometry, facilitating the construction of off-diagonal exact solutions in modified gravity theories. Finsler's geometry is fundamentally distinguished by its incorporation of the concept of anisotropy intrinsically into space-time geometry. The metric in Finsler space is expressed as a function $F(x, y)$ that maps a manifold's tangent bundle to $\mathbb{R}^1$. $F(x, y)$ is a non-negative function defined as the norm rather than the inner product that acts on a tangent bundle with the space-time coordinate $x$ and a tangent vector $y\in T_x M$ representing velocity. As a result, Finslerian geometry is dynamic, depending on both position and direction, whereas Riemannian geometry is concerned with gravitational geometry.
\par	
Recently, Z Nekouee et al. \cite{Nekouee1} explored constant-roll inflation driven by a scalar field within the Finsler model and investigated thermodynamic product formulae for Finslerian Kiselev black holes \cite{Nekouee}. J Praveen et al. \cite{Praveen} delved into the concept of cosmological constant-roll inflation within the framework of Finslerian space-time. S Banerjee et al. \cite{Sumita} examined gravastars under Finslerian space-time geometry, offering an alternative perspective to Finslerian black holes. S I Vacaru et al. \cite{Vacaru5} investigated Finsler black holes induced by Einstein gravity and studied their potential effects of quantum space-time non-commutativity. Recent research on Finsler geometry explores diverse aspects, particularly in the context of wormhole models. Rahaman et al. \cite{Rahman} presented the models of wormhole solutions in the Finslerian space-time. Manjunatha et al. \cite{Manjunatha} analyze wormhole models with an exponential shape function, incorporating anisotropic energy-momentum tensor and examining gravitational field equations in the Finslerian context.   Recently, some authors investigated the additional wormhole solutions supported by phantom energy within Finslerian geometry \cite{Singh, Rahman}. Dasa et al. \cite{Krishna} explored the potential existence of traversable WHs in the framework of Finsler–Randers (F–R) geometry. Böhmer C. G. \cite{Bohmer} derived exact solutions for traversable WHs, assuming spherical symmetry and the presence of a non-static conformal symmetry. This approach offers a systematic method for finding precise wormhole solutions.
Piyali Bhar et al. \cite{Bhar} investigated the potential existence of dark energy-supported WHs, allowing for conformal motion.
Hohmann \cite{Hohmann} explored wormhole solutions within conformal gravity. The study of WHs has extended to modified gravity theories under conformal motion, including f(G, T) gravity \cite{Mustafa, Piyali}, fourth-order conformal Weyl gravity \cite{Varieschi}, and non-commutative geometry \cite{Mustafa1}.
Hengfei Wu, in their study \cite{Hengfei}, explored traversable phantom WHs using conformal symmetry within the $f(R, \phi, \chi)$ gravity framework and a linear equation of state. In Rastall gravity, Kuhfittig et al.  \cite{Kuhfittig} found wormhole solutions with a barotropic equation of state admitting a one-parameter group of conformal motions. Banerjee et al. \cite{Banerjee} studied conformally symmetric traversable WHs in f(R, T) gravity. Recent attention has focused on WHs under conformal motion in various theories, such as f(T) gravity \cite{Ashraf}, f(Q) gravity with non-commutative geometries \cite{Rabia}, and Rastall-like-teleparallel gravity \cite{Nazavari}.
Nazavari et al. \cite{Nazavari} investigated the null energy condition at the wormhole throat, considering static and non-static conformal symmetry. Zubair \cite{Zubair} explored new Casimir WHs in f(R, T) gravity, admitting conformal Killing vectors (CKVs).
Waheed \cite{Waheed} explored obtaining analytical Yukawa–Casimir wormhole solutions in the Rastall theory of gravity using the CKVs approach.
Recently, Manjunath et al. \cite{Manju} have delved into traversable WHs in Finsler geometry under conformal motion in detail.	
\par
The paper follows a structured organization. Section \ref{sec.2} establishes the field equations utilizing Finsler geometry. Section \ref{sec.3} outlines the development of a Finslerian wormhole incorporating conformal motion. In Section \ref{sec.4}, particular solutions for the wormhole are investigated, considering a traceless fluid equation of state. Section \ref{Sec5} delves into the examination of various physical properties of Finslerian WHs, encompassing the proper radial distance, active mass, total gravitational energy, energy conditions, and the effect of anisotropy. Finally, the article concludes in Section \ref{Sec6}, offering a summary and drawing conclusions from the research.

\section{Metric formalism and Gravitational field equation in Finslerian gravity}\label{sec.2}
In Riemannian geometry, the line element is expressed as $ds=\sqrt{g_{ij}dx^idx^j}$, where $g_{ij}$ represents the metric tensor of the manifold, a function dependent on $x^i$. Finsler geometry, as a generalization of Riemannian geometry, provides a more comprehensive line element $ds^2=F^2(x^i, y^j)$. In this context, $F$ is defined on the tangent bundle of the manifold, serving as a degree 1-homogeneous function of $y^i=\frac{dx^i}{d\tau}$.
The Finsler metric $F: TM\rightarrow \mathbb{R}$, a real-valued function $F(x, y)$ at a space-time point $x$ and tangent vector $y\in T_xM$, satisfies certain conditions \cite{Bao, Shen, Bao2}.
\begin{itemize}
	\item \textbf{Regularity:} The function $F$ is infinitely differentiable ($C^\infty$) on $TM \setminus \{0\}$.
	
	\item \textbf{Positive homogeneity (of degree 1):} For any $x$ and $\kappa > 0$, we have $F(x, \kappa y) \geq \kappa F(x, y)$.
	
	\item \textbf{Strong convexity:} The matrix	$g_{ij}=\frac{1}{2}\frac{\partial^2 F^2 (x, y)}{\partial y^i \partial y^j}$ is positively defined.
\end{itemize}
The geodesic equation in Finsler space-time defines the trajectories of unbound particles affected by gravitational and related forces. It can be expressed as
\begin{equation}\label{Eq.3}
	\frac{d^{2}x^{i}}{d\tau^{2}}+2G^{i}(x, y)=0,
\end{equation}
where $G^{i}$ represents the geodesic spray coefficient, and it is of the form
\begin{equation}\label{Eq.4}
	G^{i}=\frac{1}{4}g^{ij}\left(\frac{\partial^{2}F^{2}}{\partial x^{\iota}\partial y^{j}}y^{\iota}-\frac{\partial F^{2}}{\partial x^{j}}\right).
\end{equation}
The Finslerian modified Ricci tensor is expressed as \cite{Bao, Akbar-Zadeh}
\begin{equation}\label{Eq.5}
	Ric_{ij}=\frac{\partial^{2}\bigg(\frac{1}{2}F^2 Ric\bigg)}{\partial y^{i}\partial y^{j}},
\end{equation}
The Ricci scalar $Ric$ as an invariant quantity in Finsler space-time  and is given by
\begin{equation}\label{Eq.6}
	Ric\equiv R^i_i=\frac{1}{F^2}\left (2\frac{\partial G^i}{\partial x^i}-y^\iota \frac{\partial^2 G^i}{\partial x^\iota  \partial y^i}+2G^\iota  \frac{\partial^2 G^i}{\partial y^\iota  \partial y^i}-\frac{\partial G^i}{\partial y^\iota } \frac{\partial G^\iota }{\partial y^i}\right),
\end{equation}
\begin{equation}\label{Eq.7}
	R^i_j=\frac{1}{F^2}R^i_{\iota j\sigma}y^\iota y^\sigma.
\end{equation}
The expression for $R^i_j$ involves a significant dependency on the Finsler structure $F$ and $R^i_{\iota j\sigma}$. We consider the Finsler structure $F$  in the following form \cite{Singh, Rahman, Li1}
\begin{equation}\label{Eq.8}
	F^2=e^{v(r)}y^t y^t-e^{u(r)} y^r y^r-r^2 \bar{F}^2(\theta, \phi, y^\theta, y^\theta).
\end{equation}
The given  expressions  are $v(r)=2f(r)$ and $e^{u(r)}=\left(1-\frac{S(r)}{r}\right)^{-1}$, where $f(r)$ and $S(r)$ denote the redshift and shape functions respectively. Concerning Eq. (\ref{Eq.8}), the Finsler metric and its reciprocal can be expressed as:
\begin{equation}\label{Eq.9}
	g_{ij}=diag\bigg(e^{v(r)}, -e^{u(r)}, -r^2\bar{g}_{\mu\nu}\bigg),
\end{equation}
\begin{equation}\label{Eq.10}
	g^{ij}=diag\bigg(e^{-v(r)}, -e^{-u(r)}, -r^{-2}\bar{g}^{\mu\nu}\bigg),
\end{equation}
The tensors $ \bar{g}_{\mu\nu}$ and $ \bar{g}^{\mu\nu}$ are derived from $\bar{F}$, with indices $\mu$, $\nu$ assigned to the angular coordinates $\theta$, $\phi$. As $\bar{F}$ represents the two-dimensional Finsler space \cite{Wang}, we consider $\bar{F}^2$ in the following manner:
\begin{equation}\label{Eq.11}
	\bar{F}^2=y^\theta y^\theta + \chi(\theta, \phi)y^\phi y^\phi.
\end{equation}
The Finsler metric associated with the two-dimensional Finsler structure $\bar{F}$ can be established as:
\begin{equation}\label{Eq.12}
	\bar{g}_{\mu\nu}=diag\bigg(1, \chi (\theta, \phi)\bigg),
\end{equation}
\begin{equation}\label{Eq.13}
	\bar{g}^{\mu\nu}=diag\bigg(1, \frac{1}{\chi (\theta, \phi)}\bigg).
\end{equation}
Having utilized Eq. (\ref{Eq.4}), we have successfully determined the geodesic spray coefficients for $\bar{F}^2$ as:
\begin{equation*}
	\bar{G}^\theta=-\frac{1}{4}\frac{\partial\chi}{\partial\theta}y^\phi y^\phi,
\end{equation*}
\begin{equation*}
	\bar{G}^\phi=-\frac{1}{4\chi}\left(2\frac{\partial\chi}{\partial\theta}y^\phi y^\theta + \frac{\partial\chi}{\partial\phi}y^\phi y^\phi\right).
\end{equation*}
By consulting the formula presented in the referenced source \cite{Shen, Wang}, we can ascertain the Finslerian structure's $\bar{R}ic$, denoted as $\bar{F}$, through the following procedure:
\begin{equation}\label{Eq.14}
	\bar{R}ic=\frac{1}{2\chi}\left[-\frac{\partial^2\chi}{\partial\theta^2}+\frac{1}{2\chi}\left(\frac{\partial \chi}{\partial \theta}\right)^2\right].
\end{equation}
Considering $\bar{R}ic=\lambda$ as the flag curvature, we approach the differential equation (\ref{Eq.14}) in the context of the Finsler structure $\bar{F}$ under three distinct scenarios when $(\lambda>0, \lambda=0, \lambda<0)$. Throughout our investigation, we treat $\lambda$ as a constant value, and subsequently, we obtain solutions for the differential equation in each of these situations.
\begin{equation}\label{Eq.15}
	\bar{F}^2=y^\theta y^\theta +C \sin^2(\sqrt{\lambda}\theta)y^\phi y^\phi ~~~~~~  \textrm{for} ~~ (\lambda>0),
\end{equation}
\begin{equation}\label{Eq.16}
	\bar{F}^2=y^\theta y^\theta +C \theta^2 y^\phi y^\phi  \hspace{1.6cm} ~~~ \textrm{for} ~~ (\lambda=0),
\end{equation}
\begin{equation}\label{Eq.17}
	\bar{F}^2=y^\theta y^\theta +C \sinh^2(\sqrt{-\lambda}\theta)y^\phi y^\phi ~~~ \textrm{for} ~~ (\lambda<0).
\end{equation}
When considering the solution for $\bar{F}$ with $\lambda>0$ and $C=1$ as a constant, the Finsler structure $F$ defined in Eq. (\ref{Eq.8}) transforms into the following form \cite{Singh, Rahman, Li1}
\begin{equation}\label{Eq.18}
	F^2=e^{v(r)}y^t y^t-e^{u(r)} y^r y^r-r^2 (y^\theta y^\theta+\sin^2(\sqrt{\lambda}\theta)y^\phi y^\phi).
\end{equation}
Now, we obtain the Finsler metric as
\begin{equation}\label{Eq.25}
	g_{ij}=diag\bigg(e^{v(r)}, -e^{u(r)}, -r^2, -r^2\sin^2\sqrt{\lambda}\theta\bigg),
\end{equation}
\begin{equation}\label{Eq.26}
	g^{ij}=diag\left(e^{-v(r)}, -e^{-u(r)}, -r^{-2}, (-r^2\sin^2\sqrt{\lambda}\theta)^{-1}\right).
\end{equation}
The positive flag curvature $(\lambda>0)$ significantly impacts the field equations in the context of Finslerian space-time. Utilizing Eq. (\ref{Eq.8}), we can compute the geodesic spray coefficient for the Finsler structure $F$
\begin{align*}
	G^t&=\frac{v'(r)}{2}y^t y^r,\\
	G^r&=\frac{u'(r)}{4}y^r y^r + \frac{e^{v(r)}v'(r)}{4e^{u(r)}}y^t y^t-\frac{r}{2e^{u(r)}} \bar{F}^2, \\
	G^\theta &=\frac{1}{r}y^\theta y^r+\bar{G}^\theta,\\
	G^\phi &=\frac{1}{r}y^\phi y^r+\bar{G}^\phi,
\end{align*}
where the prime $(')$ indicates the derivative with respect to $r$.
\\
The equation governing the Finslerian gravitational field is given by
\begin{equation}\label{Eq.29}
	G_{ij}=8\pi_F G T_{ij}.
\end{equation}
where
\begin{equation}\label{Eq.27}
	G_{ij}=Ric_{ij}-\frac{1}{2}g_{ij}S,
\end{equation}
\begin{equation}\label{Eq.28}
	S=g^{ij}Ric_{ij},
\end{equation}
above expression involves the Einstein tensors $G_{ij}$, scalar curvature $S$, and Ricci tensor $Ric_{ij}$. The $T_{ij}$ represents the energy-momentum tensor, $4\pi_F$ represents the volume of the two-dimensional Finsler structure $\bar{F}$, and taking $G=1$. The non-zero components of the Einstein tensor are
\begin{align}\label{Eq.30}
	G^t_t&=\frac{u'}{r e^u}-\frac{1}{r^2e^u}+\frac{\lambda}{r^2}, \nonumber\\
	G^r_r&=-\frac{v'}{re^u e^v}-\frac{1}{r^2e^u}+\frac{\lambda}{r^2}, \nonumber \\
	G^\theta_\theta &=G^\phi_\phi=-\frac{v''+{v'}^2}{2e^u}-\frac{v'}{2re^u}+\frac{u'}{2re^u}+\frac{v'}{4e^u}(u'+v').
\end{align}
Consider the general anisotropic energy-momentum tensor in the form \cite{Rahaman2010}
\begin{equation}\label{Eq.31}
	T^i_j=(\rho+p_t)\mathfrak{u}^i \mathfrak{u}_j+(p_r-p_t)\mathfrak{\eta}^i \mathfrak{\eta}_j-p_t \delta^i_j,
\end{equation}
where $\mathfrak{u}^i$ represents the four-velocity satisfying $\mathfrak{u}^i \mathfrak{u}_i =1$, while $\mathfrak{\eta}^i$ represents a space-like unit vector satisfying $\mathfrak{\eta}^i \mathfrak{\eta}_i=-1$. The terms $\rho$, $p_r$, and $p_t$ describe the energy density, radial pressure, and transverse pressure, respectively.\\
Thus, in view of Eqs. (\ref{Eq.29})-(\ref{Eq.31}), the gravitational field equations in Finsler gravity yields the following
\begin{equation}\label{Eq.32}
	8\pi_F\rho=\frac{u'}{r e^u}-\frac{1}{r^2e^u}+\frac{\lambda}{r^2},
\end{equation}
\begin{equation}\label{Eq.33}
	8\pi_Fp_r=\frac{v'}{re^u e^v}+\frac{1}{r^2e^u}-\frac{\lambda}{r^2},
\end{equation}
\begin{equation}\label{Eq.34}
	8\pi_Fp_t=\frac{v''+{v'}^2}{2e^u}+\frac{v'}{2re^u}-\frac{u'}{2re^u}-\frac{v'}{4e^u}(u'+v').
\end{equation}	
\section{Conformal symmetry in Finslerian  Wormhole}\label{sec.3}
In the WHs investigation, the conformal symmetry concept emerges as a systematic approach to seeking exact solutions. In Finsler geometry, this approach establishes a natural connection between geometry and matter through the Einstein field equations. The vector $\xi$ is identified as the generator of conformal symmetry, with the metric $g$ conformally mapped onto itself along $\xi$. In Finsler geometry,  this relationship is expressed as follows
\begin{equation}
	\mathcal{L}_\xi g_{ij}=\psi g_{ij},
\end{equation}
where $\mathcal{L}$ is the Lie derivative operator, and $\psi$ is the conformal Killing vector.
\\	
Within Finsler space-time, the metric tensor field $g$ is connected to the characterization of Conformal Killing vectors (CKVs) represented by $\xi$. The determination of these CKVs involves the application of the Lie infinitesimal operator $\mathcal{L}_{\hat{\xi}}$, leading to the following expression for $\xi$ in the Finsler space-time \cite{Joharinad}
\begin{equation}\label{Eq.1}
	\mathcal{L}_{\hat{\xi}}g_{ij}=\nabla_i\xi_j+\nabla_j\xi_i+2y^n(\nabla_n\xi^\alpha)\mathcal{C}_{\alpha ij}=\psi(r)g_{ij},
\end{equation}
where
\begin{equation}\label{Eq.71}
	\nabla_i\xi_j=\frac{\partial\xi_j}{\partial x^i}-G^\alpha_i\frac{\partial\xi_j}{\partial y^\alpha}-\Gamma^\alpha_{ij}\xi_\alpha ,
\end{equation}
and
\begin{equation}\label{Eq.72}
	\mathcal{C}_{\alpha ij}=\frac{F}{4}\frac{\partial^3F^2}{\partial y^\alpha \partial y^i \partial y^j},
\end{equation}
Here, the symbol $\mathcal{L}$ denotes the Lie derivative operator, and $\psi$ represents the conformal factor. The CKVs are expressed as $\hat{\xi}=\xi^i(r)\frac{\partial}{\partial x^i}+y^j(\partial_j\xi^i)\frac{\partial}{ \partial y^i}$. The Cartan connection in the context of the Finslerian wormhole structure, denoted by $\mathcal{C}_{\alpha ij}$ in Eq. (\ref{Eq.8}), is subject to the observation that when $\mathcal{C}_{\alpha ij}$ equals zero for all indices $\alpha$,  $i$, and  $j$, this condition characterizes the geometric properties of the Finslerian wormhole space-time.

The Cartan connection in the Finslerian wormhole space-time is established through the CKVs $\xi$, which introduces conformal symmetry. The outcomes derived from Eq. (\ref{Eq.1}) are contingent on the specific value assigned to $\psi$. For instance, if $\psi=0$, it generates a Killing vector, indicating the asymptotic flatness of space-time. According to discussions in \cite{Herrera1}, if $\psi=\psi(x, t)$, it leads to conformal vectors, while a constant value of $\psi$ results in a homothetic vector. By introducing $\xi^{i}=g^{ij}\xi_j$, the expression derived from Eq. (\ref{Eq.1}) are as follows \cite{Bohmer}
\begin{align*}
	\xi^r v'&=\psi,\\
	\xi^t&=a_1,\\
	\xi^r&=\frac{\psi r}{2},\\
	\psi^ru'+2\psi^r_{,r}&=\psi.
\end{align*}
In our current study, the symbol $r$ signifies spatial coordinates and $t$ represents temporal coordinates. Specifically, the conformal factor $\psi$ is considered a function of the radial coordinate $r$.
\begin{equation}\label{Eq.36}
	e^v=e^{2f(r)}=a_2^2r^2,
\end{equation}
\begin{equation}\label{Eq.37}
	e^u=\left(1-\frac{S(r)}{r}\right)^{-1}=\left(\frac{a_3}{\psi}\right)^2,
\end{equation}
\begin{equation}\label{Eq.38}
	\xi^i=a_1\delta^i_t+\left(\frac{\psi r}{2}\right)\delta^i_r,
\end{equation}
By applying Eqs. (\ref{Eq.36}-\ref{Eq.38}), the field equations (\ref{Eq.32}-\ref{Eq.34}) can be reformulated with the integration constants $a_1$, $a_2$, and $a_3$. So, we have
\begin{equation}\label{Eq.39}
	8\pi_F\rho=\frac{1}{r^2}\left[\lambda-\frac{\psi^2}{a^2_3}\right]-\frac{2\psi'\psi}{ra_3^2},
\end{equation}
\begin{equation}\label{Eq.40}
	-8\pi_F p_r=\frac{1}{r^2}\left[\lambda-\frac{3\psi^2}{a^2_3}\right],
\end{equation}
\begin{equation}\label{Eq.41}
	8\pi_F p_t=\frac{\psi^2}{r^2a^2_3}+\frac{2\psi \psi'}{r a^2_3}.
\end{equation}
\section{Finslerian Wormholes solution with Traceless fluid}\label{sec.4}
In this section, we explore a potential solution for the Finslerian wormhole by considering an alternative equation of state (EoS) known as the traceless fluid \cite{Saiedi, Sharif1}
\begin{align}\label{Eq.54}
	T=0~~~~~\Rightarrow ~~~~~\rho-p_r-2p_t=0,
\end{align}
where $T = g^{ij} T_{ij}$, the traceless energy-momentum tensor holds significance in the Finslerian WHs study because it plays a crucial role in formulating the gravitational field equations within Finsler space-time. By using Eqs. (\ref{Eq.39}-\ref{Eq.41}) in Eq. (\ref{Eq.54}), we obtain conformal factor
\begin{equation}\label{Eq.55}
	\psi^2(r)=\frac{a_3^2\lambda r^2+3b_1}{3r^2},
\end{equation}
where $b_1$ is an integration constant. Using Eq. (\ref{Eq.55}) we can obtain the shape function $S(r)$ from Eq. (\ref{Eq.37})
\begin{equation}\label{Eq.56}
	S(r)=\frac{-r^2(\lambda-3)a_3^2-3b_1}{3ra_3^2}.
\end{equation}
In view of  Eq. (\ref{Eq.55}), the field equations (\ref{Eq.39}-\ref{Eq.41}) under conforam motion have the following form
\begin{equation}\label{Eq.57}
	\rho=\frac{2a_3^2\lambda r^2+3b_1}{24r^4a_3^2\pi},
\end{equation}
\begin{equation}\label{Eq.58}
	p_r=\frac{3b_1}{r^4a_3^2 8\pi},
\end{equation}
\begin{equation}\label{Eq.59}
	p_t=\frac{a_3^2\lambda r^2-3b_1}{24r^4a_3^2\pi},
\end{equation}
\begin{equation}\label{Eq.60}
	\rho+p_r=\frac{2}{3}\left[ \frac{a_3^2\lambda r^2+6b_1}{3r^4a_3^2 8\pi}\right],
\end{equation}
\begin{equation}\label{Eq.62}
	\rho+p_t=\frac{\lambda}{r^28\pi},
\end{equation}
\begin{equation}\label{Eq.61}
	\rho-p_r=\frac{2}{3}\left[ \frac{a_3^2\lambda r^2-3b_1}{r^4a_3^28\pi}\right],
\end{equation}

\begin{equation}\label{Eq.63}
	\rho-p_t=\frac{1}{3}\left[ \frac{a_3^2\lambda r^2+6b_1}{r^4a_3^2 8\pi}\right],
\end{equation}
\begin{equation}\label{Eq.64}
	p_t-p_r=\frac{1}{3}\left[ \frac{a_3^2\lambda r^2-12b_1}{r^4 a_3^2 8\pi}\right],
\end{equation}
\begin{equation}\label{Eq.65}
	\rho+p_r+2p_t=\frac{1}{3}\left[ \frac{4a_3^2\lambda r^2+6b_1}{r^4a_3^2 8\pi}\right].
\end{equation}
\subsection{The shape function and redshift function}\label{4.1}	
\begin{figure}[hptb]
	\begin{center}
		{\includegraphics[scale=0.3]{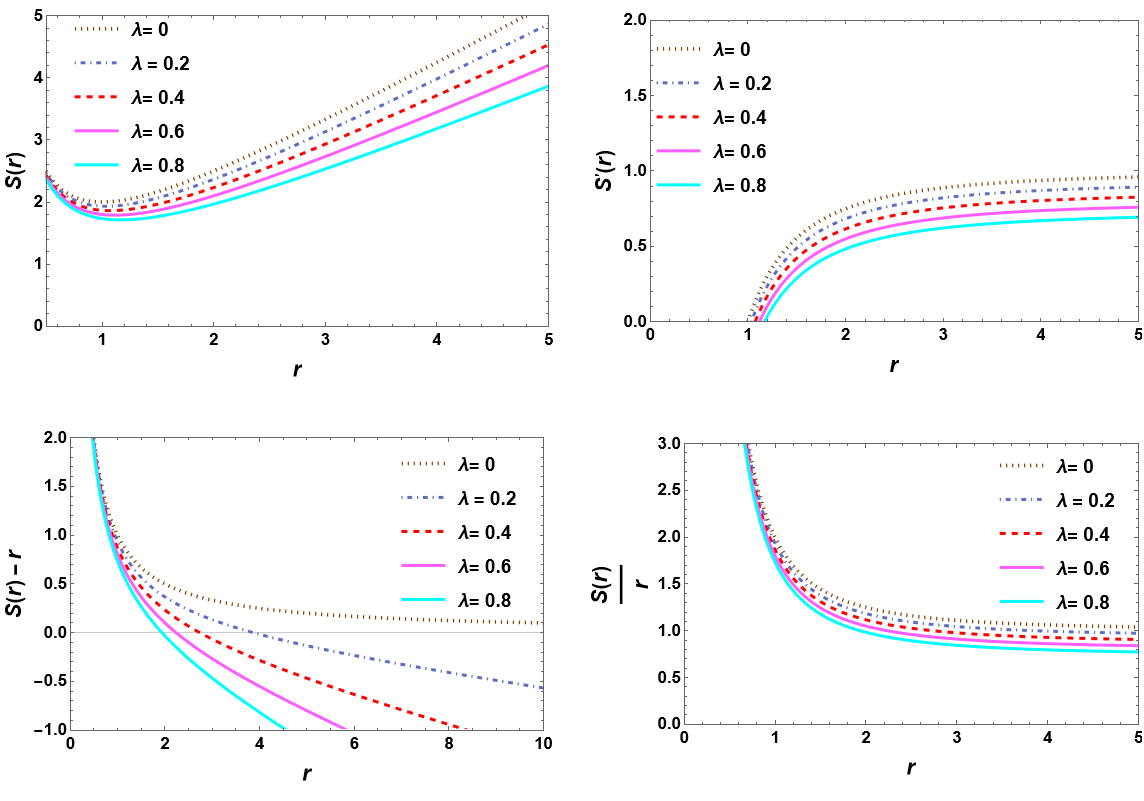}}
		\caption{\label{fig1} The figures illustrate the behavior of the shape function described by Eq.(\ref{Eq.56}) for different values of $\lambda$ in table-\ref{tbl1},  keeping the value of constant $b_1 = -1$ and $a_3 = 1$.}
	\end{center}
\end{figure}

In our analysis, we consider the functions $v(r) = f(r)$ and $e^{u(r)} = \left(1-\frac{S(r)}{r}\right)^{-1}$ from the Finslerian metric Eq. (\ref{Eq.18}), where $f(r)$ and $S(r)$ are denoted as the redshift function and shape function, respectively. The shape function, describing the geometric features of space-time curvature in the region of these theoretical structures, plays a crucial role in the traversable WHs analysis.
For these functions to meet specific conditions, they should be expressed as arbitrary functions of the radial coordinate $r$. Let's delve into the shape function derived under conformal motion, as discussed in \cite{Morris}.
\begin{itemize}
	\item The wormhole throat is the connection between two asymptotic regions and is situated at the radial coordinate $r_0$, characterized by the condition $S(r_0) = r_0$.
	\item The flaring-out condition, given by $\frac{S(r)-rS'(r)}{2S^2(r)}> 0$ and applicable at or near the throat, must be fulfilled by the shape function $S(r)$. This requirement further simplifies to $S'(r_0) < 1$ in the close vicinity of the wormhole's throat.
	\item To maintain the proper signature of the metric for radial coordinates $(r > r_0)$, the shape function must fulfill the condition $(1 - \frac{S(r)}{r} > 0)$.
	\item For the metric functions to exhibit asymptotically flat geometries, they must satisfy the conditions that $f(r)$ and $\frac{S(r)}{r}\rightarrow 0$  as $r\rightarrow\infty$. It's important to note that these requirements can be relaxed for non-asymptotically flat WHs.
	\item Additionally, it is crucial for the function $f(r)$ to be finite and nonzero across the entire space-time to prevent the presence of horizons and singularities.
\end{itemize}
Geometric characteristics of the shape function Eq. (\ref{Eq.56}), as illustrated in Fig. (\ref{fig1}) for various values of $\lambda$, depict $S(r)$, $S'(r)$, $S(r)-r$ and $S(r)/r$. The point at which $S(r)-r$ intersects the $r$-axis indicates the throat of the wormhole located at $r = r_0$. The positions of the wormhole throat for various values of $\lambda$ are presented in Table. (\ref{tbl1}).
In Fig. (\ref{fig1}), it is evident that $S'(r) < 1$, adhering to the flaring-out requirement. Moreover, as illustrated in the figure, the asymptotic behavior of $S(r)/r$ does not tend to zero as $r$ approaches $\infty$. The redshift function $f(r)$ must be finite across the entire space-time to prevent the emergence of horizons within the wormhole. The requirement for the finiteness of $f(r)$ is crucial to maintain the wormhole traversability and to avoid the occurrence of singularities or event horizons \cite{Krishna, Konoplya}. This behavior in our study suggests that the wormhole space-time does not exhibit asymptotically flat characteristics \cite{Piyali}.

\begin{table}[htbp]
	\centering
	\caption{The position of wormhole throat $r_0$ under different values of $\lambda$.}\label{tbl1}
	\begin{tabular}{lcc}
		\hline
		\bf{${\bf\lambda}$} & \bf{$r_0$}  \\
		\hline
		0  & -      \\	
		0.2& 3.8729 \\	
		0.4& 2.7386 \\	
		0.6& 2.2361 \\
		0.8& 1.9365 \\
		\hline	
	\end{tabular}
\end{table}
\subsection{Embedding surface}\label{4.2}
\begin{figure}[hptb]
	\begin{center}
		{\includegraphics[scale=0.41]{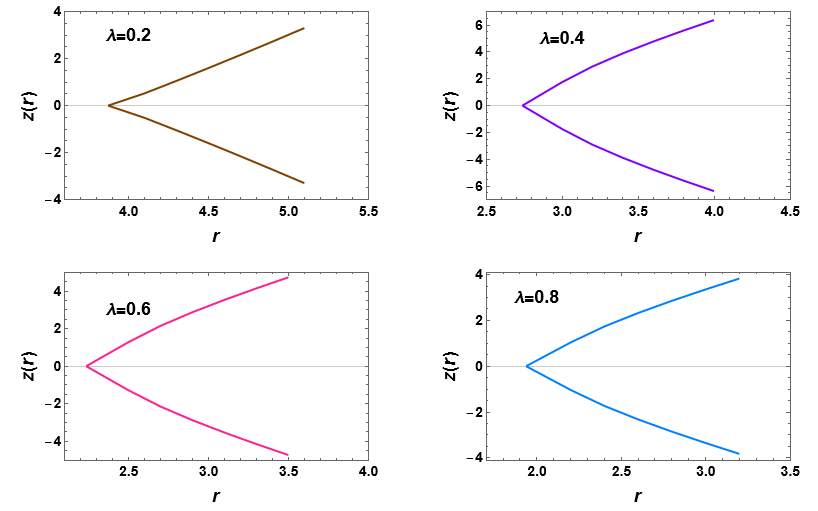}}
		\caption{\label{fig2} An embedding 2-D diagram of Finslerian WHs with different $\lambda$ values.}
	\end{center}
\end{figure}

\begin{figure}[hptb]
	\begin{center}
		{\includegraphics[scale=0.41]{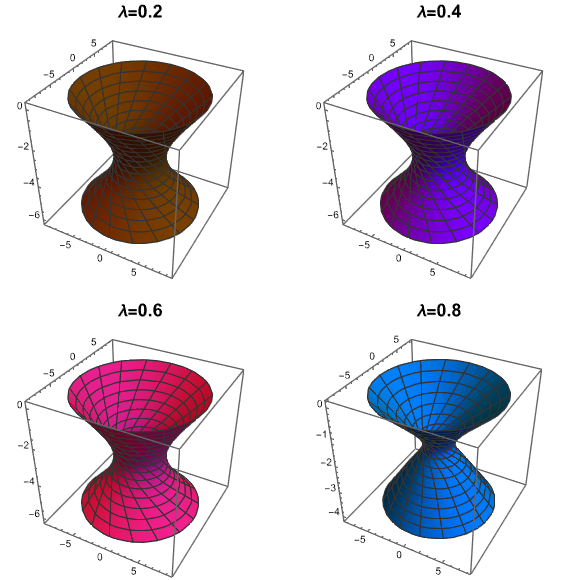}}
		\caption{\label{fig3} 3-D representations of Finslerian WHs, each corresponding to different $\lambda$ values.}
	\end{center}
\end{figure}

For a clearer visualization of the wormhole geometry described by Eq. (\ref{Eq.18}), we have created embedded 2-D and 3-D diagrams illustrating the shape function from Eq. (\ref{Eq.56}). Due to the spherically symmetric nature of our wormhole geometry, we opted for an equatorial slice with $(\theta = \frac{\pi}{2})$ and fixed a specific moment in time with $t =$constant, chosen without loss of generality. This chosen hypersurface is represented as $h: \theta = \frac{\pi}{2}, t = \text{constant}$. Under this condition, $d\theta = 0$ and $dt = 0$. As a result, the reduced form of the wormhole metric (\ref{Eq.18}) can be written as
\begin{equation}\label{Eq.49}
	F^2_h=-\frac{dr^2}{1-\frac{S(r)}{r}}-r^2d\phi^2.
\end{equation}
The metric Eq. (\ref{Eq.49}) has the following equivalent form in cylindrical coordinates $(r,z,\phi)$
\begin{equation}\label{Eq.50}
	F^2_h=-dZ^2-dr^2-r^2d\phi^2.
\end{equation}
The embedding surface in Euclidean three-dimensional space is represented by $Z = Z(r)$. Consequently, the metric for this surface in reduced form can be expressed as \cite{Lobo}
\begin{equation}\label{Eq.51}
	F^2_h=-\left[1+\left(\frac{dZ}{dr}\right)^2\right]dr^2-r^2d\phi^2.
\end{equation}
We can express the embedding shape function $Z(r)$ by comparing Eqs. (\ref{Eq.49}) and (\ref{Eq.50}) as follows
\begin{equation}\label{Eq.52}
	\frac{dZ}{dr}=\pm\sqrt{\left(1-\frac{S(r)}{r}\right)^{-1}-1}.
\end{equation}
The differential equation (\ref{Eq.52}) shows divergence at the wormhole throat, indicating that the embedding surface turns vertical. Eq. (\ref{Eq.50}) provides the following expression
\begin{equation}\label{Eq.53}
	Z(r)=\pm\int_{r_0^+}^{r}\sqrt{\left(1-\frac{S(r)}{r}\right)^{-1}-1}.
\end{equation}
Since the analytical method does not exist to solve the integral in Eq. (\ref{Eq.53}), we can use numerical solutions. The numerical values are obtained by varying the upper limit of $r$ and fixing a particular lower limit value of $r^0_+$ outside the wormhole throat, results shown in Tables (\ref{tbl2}-\ref{tbl5}) for various values of $\lambda$.
The embedding diagram $Z(r)$ of WHs with different values of $\lambda$ is shown in Fig. (\ref{fig2}). Revolving Fig. (\ref{fig2}) around the 'z-axis' yields Fig. (\ref{fig3}), which presents the comprehensive visualization of the WHs for various $\lambda$ values as shown in the figures.

\begin{table}[htbp]
	\centering
	\caption{For various $r$, the values of $Z(r)$, $l(r)$, and $M(r)$ are obtained at fixed $r_0^+ = 3.9$, $\lambda=0.2$, with $a_3=1$ and $b_1=-1$.}\label{tbl2}
	\begin{tabular}{lcccc}
		\hline
		\bf{${r}$} & \bf{$Z(r)$} & l(r) & M(r) & $E_{Tg}$  \\
		\hline
		4.1&	0.51925	&   3.44792&	0.01292&1.9364797
		\\
		4.3&	1.05237	&   5.52916&	0.02588&1.9365106
		\\
		4.5&	1.59829&	7.26383&	0.03887&1.9365422
		\\
		4.7&	2.15602&	8.82794&	0.05189&1.9365743
		\\
		4.9&	2.72467&	10.28924&	0.06494&1.9366066
		\\
		5.1&	3.30343&	11.68163&	0.07801&1.9366389
		\\
		\hline	
	\end{tabular}
\end{table}

\begin{table}[htbp]
	\centering
	\caption{For various $r$, the values of $Z(r)$, $l(r)$, and $M(r)$ are obtained at fixed $r_0^+ = 2.8$, $\lambda=0.4$, with $a_3=1$ and $b_1=-1$.}\label{tbl3}
	\begin{tabular}{lcccc}
		\hline
		\bf{${r}$} & \bf{$Z(r)$}&l(r)&M(r)&$E_{Tg}$   \\
		\hline
		3   &   1.74632&	1.75814&	0.02514&1.3693808
		\\
		3.2	&   2.91499&	2.94397&	0.05048&1.3694652
		\\
		3.4 &	3.89436&	3.94387&	0.07598&1.3695514
		\\
		3.6 &	4.77545&	4.84795&	0.10160&1.3696383
		\\
		3.8 &	5.59618&	5.69355&	0.12733&1.3697251
		\\
		4  &  	6.37623&	6.49998&	0.15316&1.3698113
		\\
		\hline	
	\end{tabular}
\end{table}

\begin{table}[htbp]
	\centering
	\caption{For various $r$, the values of $Z(r)$, $l(r)$, and $M(r)$ are obtained at fixed $r_0^+ = 2.3$, $\lambda=0.6$, with $a_3=1$ and $b_1=-1$.}\label{tbl4}
	\begin{tabular}{lcccc}
		\hline
		\bf{${r}$} & \bf{$Z(r)$}&l(r)&M(r)&$E_{Tg}$   \\
		\hline
		2.5&	1.28040&	1.29644&	0.03671&1.1181937
		\\
		2.7&	2.14568&	2.18474&	0.07391&1.1183442
		\\
		2.9&	2.87391&	2.94032&	0.11149&1.1184976
		\\
		3.1&	3.53117&	3.62802&	0.14939&1.1186515
		\\
		3.3&	4.14508&	4.27470&	0.18754&1.1188044
		\\
		3.5&	4.72995& 	4.89417&	0.22590&1.1189553
		\\
		\hline	
	\end{tabular}
\end{table}

\begin{table}[htbp]
	\centering
	\caption{For various $r$, the values of $Z(r)$, $l(r)$, and $M(r)$ are obtained at fixed $r_0^+ = 2$, $\lambda=0.8$, with $a_3=1$ and $b_1=-1$.}\label{tbl5}
	\begin{tabular}{lcccc}
		\hline
		\bf{${r}$} & \bf{$Z(r)$}&l(r)&M(r)&$E_{Tg}$  \\
		\hline
		2.2&	1.03421&	1.05400&	0.04765&0.9684654
		\\
		2.4&	1.73331&	1.78139&	0.09624&0.9686914\\
		2.6&	2.32236&	2.40394&	0.14557&0.9689212
		\\
		2.8&	2.85473&	2.97344&	0.19547&0.9691509
		\\
		3  &    3.35262&	3.51121&	0.24583&0.9693778\\
		3.2&	3.82755&	4.02813&	0.29656&0.9696006
		\\
		\hline	
	\end{tabular}
\end{table}

\section{Physical viability of Finslerian Wormhole model}\label{Sec5}
The section focuses on scrutinizing the physical characteristics of the proposed Finslerian wormhole model when subjected to conformal motion in the presence of a traceless fluid. The investigation aims to understand how the wormhole's essential features respond to the geometric conformal transformation when considering the fluid with a traceless energy-momentum tensor. This analysis contributes to verifying the model's robustness and applicability in scenarios where conformal symmetry and traceless fluids play significant roles in theoretical physics.

\subsection{Proper radial distance in Finslerian wormholes}\label{sec5.1}
\begin{figure}[hptb]
	\begin{center}
		{\includegraphics[scale=0.35]{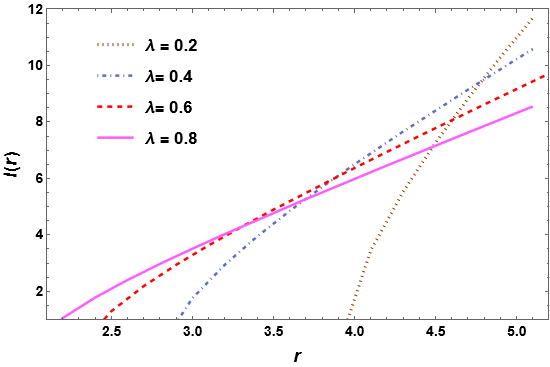}}
		\caption{\label{fig4} The plot depicts the proper radial distance v/s $r$ for different $\lambda$ values, reflecting the characteristics presented in Fig. (\ref{fig1}).}
	\end{center}
\end{figure}

The wormhole's proper radial distance is determined by the following \cite{Krishna, Banerjee}
\begin{equation}
	l(r)=\pm\int_{r_0^+}^{r}\frac{dr}{1-\frac{S(r)}{r}}.
\end{equation}
The proper radial distance integral has been evaluated over a spectrum of values through numerical integration, and the findings are detailed in Tables (\ref{tbl2}-\ref{tbl5}). Fig. (\ref{fig4}) depicts the variations in the profiles of $l(r)$ corresponding to different values of $\lambda$ in Table (\ref{tbl1}).
\subsection{Active mass profiles of  Finslerian wormholes}\label{sec5.2}
\begin{figure}[hptb]
	\begin{center}
		{\includegraphics[scale=0.4]{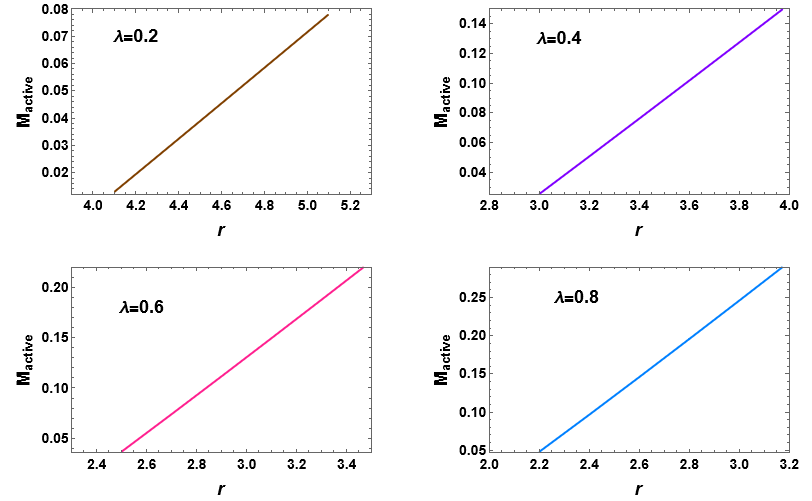}}
		\caption{\label{fig5} The active mass function profile of the Finslerian wormhole for various $\lambda$ values.}
	\end{center}
\end{figure}

In general relativity, the active mass function is associated with the distribution of matter and energy in a given space-time region, influencing the gravitational effects observed in that region. This is a fundamental component in understanding how matter and energy contribute to the curvature of space-time and the resulting gravitational interactions \cite{Piyali1}. The active mass function $M_{act}$ is defined by
\begin{equation}\label{Eq.59a}
	M_{act}=\int_{r_0^+}^{r} 4\pi\rho r^2 dr
\end{equation}
Using Eq. (\ref{Eq.57}) in above, we obtain
\begin{equation}\label{Eq.59a}
	M_{act} =\left[\frac{2a_3^2\lambda r^2-3b_1}{6a_3^2r}\right]^r_{r_0^+}
\end{equation}
Fig. (\ref{fig5}) illustrates the active mass function graph for the Finslerian wormhole, displaying changes with different values of  $\lambda$. Notably, $M_{act}$ exhibits a consistent positive trend and monotonic increase with $r$, particularly outside the wormhole throat, as observed in Fig. (\ref{fig5}).
\subsection{Total gravitational energy of Finserian wormholes}
\begin{figure}[hptb]
	\begin{center}
		{\includegraphics[scale=0.39]{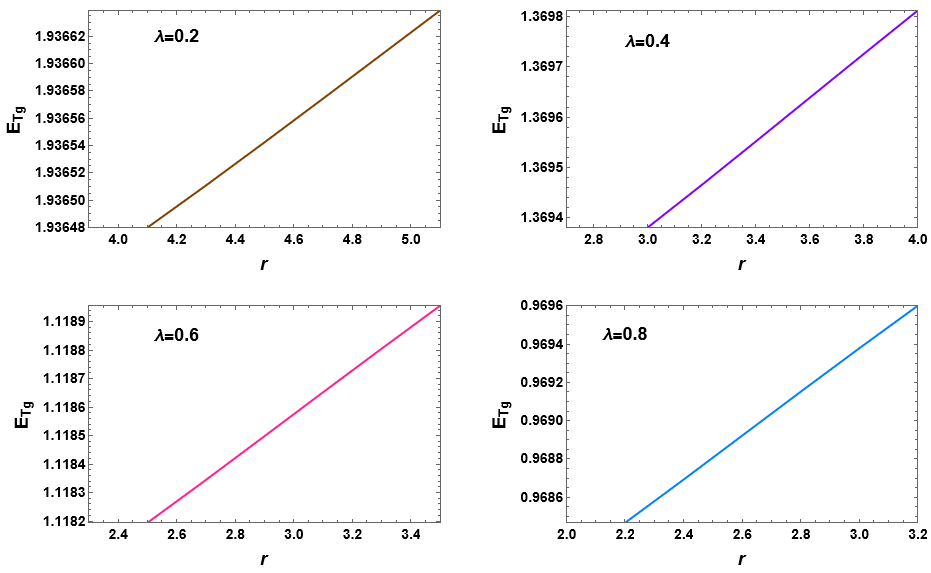}}
		\caption{\label{fig6} The presentation of the total gravitational energy of the Finslerian WHs is depicted for various values of $\lambda$.}
	\end{center}
\end{figure}

The total gravitational energy ($E_{Tg}$) of a Finslerian wormhole is an important aspect in the study of these exotic structures within the framework of Finsler geometry under conformal symmetry. This quantity encapsulates the gravitational effects associated with the wormhole's geometry and mass distribution. The $E_{Tg}$ can be defined as \cite{Lynden, Nandi}
\begin{equation}\label{Eq.61a}
	E_{Tg}=Mc^2-E_M.
\end{equation}
The quantities $Mc^2$ and $E_M$, referred to as the total energy and total mechanical energy, can be expressed as
\begin{equation}\label{Eq.62a}
	Mc^2=\frac{1}{2}\int_{r_0^+}^{R}\rho r^2 dr+\frac{r_0}{2},
\end{equation}
\begin{equation}\label{Eq.63a}
	E_M=\frac{1}{2}\int_{r_0^+}^{r}\sqrt{g_{rr}}\rho r^2 dr.
\end{equation}
Upon substituting Eqs. (\ref{Eq.62a}) and (\ref{Eq.63a}) into Eq. (\ref{Eq.61a}), the resulting expression yields the $E_{Tg}$ as follows,
\begin{equation}
	E_{Tg}=\frac{1}{2}\int_{r_0^+}^{r}\left(1-\sqrt{g_{rr}}\right)\rho r^2 dr+\frac{r_0}{2},
\end{equation}
where $\sqrt{g_{rr}}=\left( 1-\frac{S(r)}{r}\right)^{-1}$. We conducted numerical integration for different values of $\lambda$, with the values given in Tables  (\ref{tbl2}-\ref{tbl5}). Fig. (\ref{fig6}) graphically represents the profiles of $E_{Tg}$ with $r$ associated with different $\lambda$ values.
\subsection{Energy conditions}
This section discusses the physical characteristics of our WHs model, analyzing various energy conditions (EC) because astrophysical events are significantly influenced by such dynamic conditions. The energy conditions lie in the geometric interplay between the Raychaudhuri equation and gravitational field and are used to confirm how energy conditions behave in any gravitational framework \cite{Raychaudri, Nojiri}. These conditions yield numerous valuable and general results that find applicability across a diverse spectrum of cosmological phenomena. The exploration of various energy scenarios and the methodology for predicting the existence of WHs is a crucial aspect of our study. Violations of energy criteria serve as significant indicators for forecasting the presence of WHs, providing valuable insights into the geometry of traversable WHs \cite{ Manjunatha, Banerjee, Nazavari}
\par
\begin{figure}[hptb]
	\begin{center}
		{\includegraphics[scale=0.41]{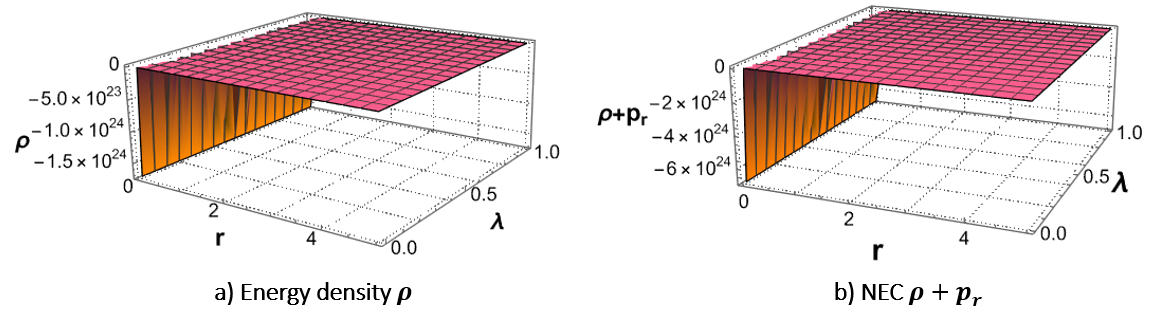}}
		{\includegraphics[scale=0.41]{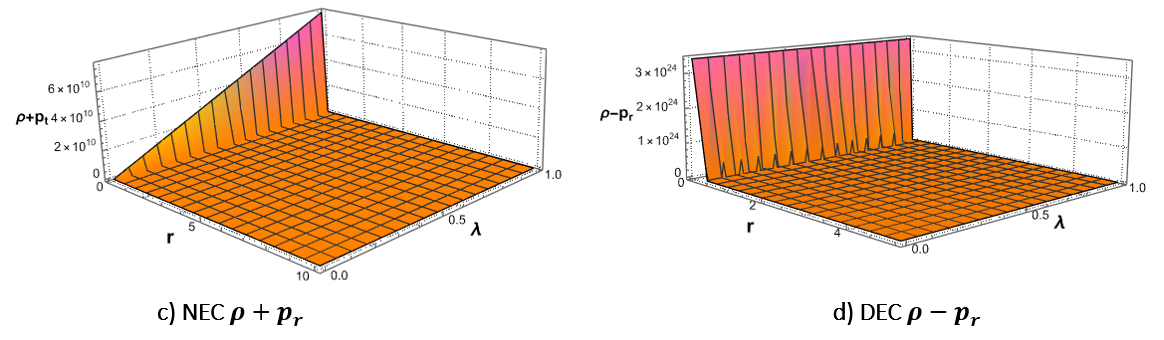}}
		{\includegraphics[scale=0.41]{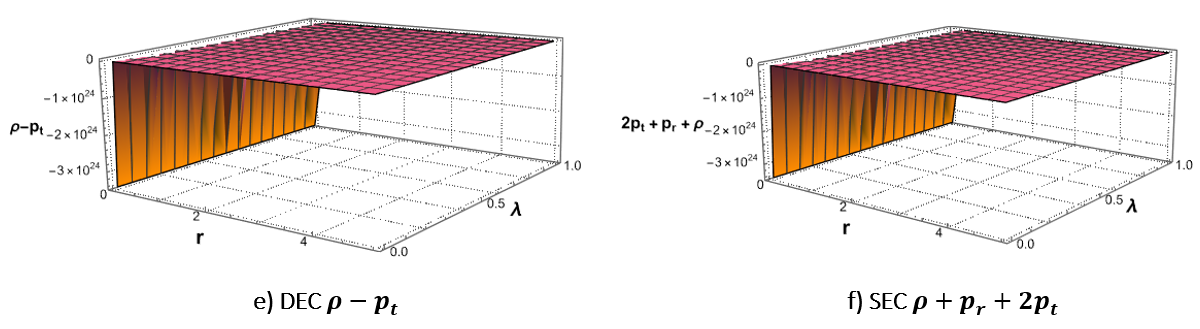}}
		\caption{\label{fig7}Profile of energy conditions of Finslerian wormhole for different values of $\lambda$ with $a_3=1$ and $b_1=-1$.}
	\end{center}
\end{figure}

The Physical quantities $\rho$, $p_r$, and $p_t$  play a fundamental role in defining the energy conditions, encompassing the null energy condition (NEC), dominant energy condition (DEC), weak energy condition (WEC), and strong energy condition (SEC) are \cite{Morris, Morris1988}
\begin{itemize}
	\item [*] ${\bf NEC:} \rho+p_x\geq 0$, ~~$ \forall ~~x$,
	\item [*] ${\bf DEC:}  \rho\geq 0$ and $\rho-p_x\geq0 ~~\forall ~~x$,
	\item [*] ${\bf WEC:}  \rho\geq 0$ and $\rho+p_x \geq0 ~~\forall ~~x$,
	\item [*] ${\bf SEC:}  \rho+p_x\geq 0$ and $\rho+\sum_x p_x\geq 0 ~~\forall~~ x$.~~~~~~~~~~~where $x=t, r.$
\end{itemize}
The observed violations of the NEC and WEC are indicative of violations in the remaining two constraints.
Notably, the NEC violation establishes a mandatory criterion for the existence of wormhole geometry.
Our investigation focuses significantly on understanding how the parameter $\lambda$ in the Finsler geometry affects the violation of energy conditions. For further understanding, we systematically analyzed the energy conditions and visually represented their patterns through graphical visualization in Fig. (\ref{fig7}).

\begin{itemize}
	\item In the traceless fluid model, NEC is violated for all the values of  $\lambda\geq0$ against the wormhole radius $r$ expressed by the condition $\rho+p_r<0$ as depicted in Fig. (\ref{fig7}b) and is valid in terms of $\rho+p_t\geq0$ shown in Fig. (\ref{fig7}c).
	\item In our specific model based on our observations from Fig. (\ref{fig7}d) and Fig. (\ref{fig7}e), we find that the DEC is violated in terms of $\rho-p_t<0$ and valid in terms of$\rho-p_r>0$ respectively for all values of $\lambda\geq0$.
	Further, Fig. (\ref{fig7}a) illustrates that $\rho<0$ is violated for all values of $\lambda\geq0$ in the vicinity of the wormhole throat.
	\item  WEC is violated in terms of $\rho<0$ see from Fig. (\ref{fig7}a) and $\rho+p_r<0$ from Fig. (\ref{fig7}b) for all the values of $\lambda\geq0$ and is valid in terms of $\rho+p_t\geq0$ against the values of $\lambda\geq0$ near the wormhole throat.
	\item The impressive point is that SEC is also violated in terms of $\rho+p_r+2p_t<0$ in Fig. (\ref{fig7}f) for all the values of $\lambda\geq0$  near the wormhole throat.
\end{itemize}
\subsection{Effect of anisotropy}
\begin{figure}[hptb]
	\begin{center}
		{\includegraphics[scale=0.32]{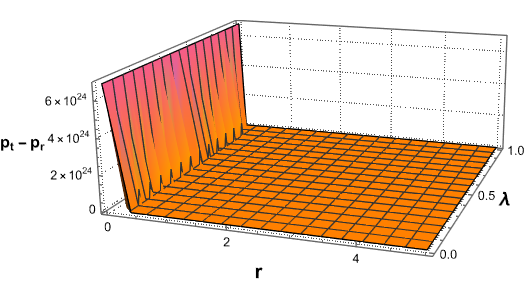}}
		\caption{\label{fig8} Profile of anisotropy of Finslerian WHs for various values of $\lambda$ with $a_3=1$ and $b_1=-1$.}
	\end{center}
\end{figure}

Anisotropy analysis is very important because it reveals information about the internal structure of a relativistic wormhole configuration and is used to quantify anisotropy within the wormhole \cite{Rahaman1, Sharif, Shamir}. The anisotropic parameter is defined by
\begin{equation}
	\Delta=p_t-p_r.
\end{equation}
Within the context of WHs, the parameter $\Delta$ serves as a measure of anisotropy, commonly referred to as the anisotropy factor. A positive value of $\Delta$ characterizes the geometry of WHs as repulsive, while a negative $\Delta$ implies an attractive geometry. When $\Delta = 0$, the matter distribution within the WHs exhibits isotropic pressure, signifying that the pressures in the radial and tangential directions are equal. In our investigation, the anisotropy factor ($\Delta$) is depicted as positive for the entire range $\lambda \geq 0$, as illustrated in Fig. (\ref{fig8}).
\section {Summary and conclusion}\label{Sec6}
In this research, we expandly have explored wormhole solutions within the framework of Finsler geometry by integrating conformal symmetries with traceless fluid. The viability of traversable Finslerian WHs has been meticulously analyzed across various scenarios.
Finsler geometry has provided a new perspective on manifold structure and reveals a profound influence on wormhole physics. We derived the field equations specific to Finslerian WHs under a conformal motion scenario by employing a traceless  EoS model. We extracted the conformal factor $\psi(r)$. The elucidation of conformal motion and metric transformations in Finslerian wormhole solutions has been significantly shaped by the essential contributions of the conformal factor ($\psi(r)$) and CKVs. To achieve this goal, we studied some physical aspects of WHs. The key aspects of the current study are summarized as follows:
\begin{itemize}
	\item We have successfully derived the shape function $S(r)$ and conducted a detailed examination of their dependence on the Finslerian parameter $\lambda$. The resulting shape function adheres to all essential criteria, as depicted in Fig. (\ref{fig1}).
	
	\item Utilizing the derived shape function, we present the wormhole throat radius for various values of $\lambda$ in Table  (\ref{tbl1}). Based on these results, we visualize the 2D and 3D embedding diagrams of Finslerian WHs, as shown in Figs. (\ref{fig2}) and (\ref{fig3}), respectively, with different values of $\lambda$.
	
	\item To gain a comprehensive understanding of the traversable Finslerian wormhole, we focused on studying the proper radial distance $l(r)$. This property is a crucial tool for detecting singularities and avoiding space-time horizons. In our study, $l(r)$ demonstrates a consistently increasing behavior for different values of $\lambda$ shown in Fig. (\ref{fig4}). We note that the active mass function remains positive across various $\lambda$ values, as illustrated in Fig. (\ref{fig5}).
	
	\item Another crucial physical property affirming the traversability of the wormhole is the total gravitational energy $(E_{Tg})$. If $(E_{Tg})$ is too strong, particles may not be able to pass through the throat, leading to black hole formation. In our study, $(E_{Tg})$ exhibits lower values near the throat and increases towards the wormhole surface shown in Fig. (\ref{fig5}) for different values of $\lambda$. This observation strongly supports the traversability of our proposed Finslerian wormhole under conformal motion.
	
	\item To demonstrate the existence of exotic matter, such as traceless fluid, we examined the energy conditions illustrated in Fig. (\ref{fig7}). In our model, all energy conditions violation, including NEC, WEC, DEC, and SEC, confirms the presence of exotic matter near the wormhole throat. Similar violations of energy conditions have been studied in \cite{Singh, Shahzad}, providing valuable insights. Additionally,  Fig. (\ref{fig8}) shows that the anisotropic factor is positive, indicating a repulsive nature in the wormhole geometry. This conclusion supports the traversability of the Finslerian wormhole under conformal motion.
\end{itemize}
The findings suggest that Finslerian WHs are physically achievable creatures. The analysis of conformal transformations on light ray trajectories has potential applications in observational astrophysics, providing new opportunities to detect and study these exotic objects. Additionally, exploring the effects of conformal transformations on the quantum properties of space-time near Finslerian WHs can contribute to the quantum gravity theories improvement.

\section*{Author contributions}
Manjunath Malligawad wrote the main manuscript text and plot the all graphs. Z. Nekouee contributed to the formulation. All authors reviewed the manuscript.

\section*{Declaration of competing interest}
The authors affirm that they do not have any known competing financial interests or personal relationships that could have been perceived to influence the work reported in this paper.

\section*{Data availability}
No data associated in the manuscript.

\section*{Acknowledgment}
The author Manjunath Malligawad acknowledges OBC Fellowship, Backward Classes Welfare Department (BCWD), Govt. of Karnataka, India, for the financial support in the form of OBC-Ph.D Fellowship (Application No: 2021PHD9521296). The author RK is thankful to IUCAA for its visiting associateship program and the part of this work was done during his visit in IUCAA, Pune.


\begin{thebibliography}{45}
	
	\bibitem{Einstein}
	A. Einstein,  N. Rosen, The particle problem in the general theory of relativity, \textit{Phys. Rev.} \textbf{48} (1935) 73-77.
	
	\bibitem{Morris}
	M.S. Morris, K.S. Thorne, Wormholes in spacetime and their use for interstellar travel: A tool for teaching general relativity, \textit{Am. J. Phys.} \textbf{56} (1988) 395-412.
	
	
	\bibitem{Weyl}
	H. Weyl, Feld und materie, \textit{Annalen der Physik} \textbf{370} (1921) 541.
	
	\bibitem{Saaidi}
	K. Saaidi,  S. Tavakoli, Wormholes supported by dark energy in f(T) gravity,  \textit{Phys. Dark Universe}  \textbf{31} (2021) 100763.
	
	\bibitem{Singh}
	K. Singh, F. Rahaman, D. Deb, S.K. Maurya, Traversable Finslerian wormholes supported by phantom energy,  \textit{Front. Phys.}  \textbf{10} (2023) 1038905.
	
	\bibitem{Bao}
	D. Bao, S.S. Chern, Z. Shen, {An Introduction to Riemann Finsler Geometry}, Graduate Texts in Mathematics 200, Springer, New York, 2000.
	
	\bibitem{Cartan}
	E. Cartan,   Les Espaces de Finsler, Actualite Scientifiques et Industrielles, Paris, Hermann, 1934.
	
	\bibitem{Horvath}
	J.I. Horvath, A Geometrical Model for the Unified Theory of Physical Fields,  \textit{Phys. Rev.}  \textbf{80} (1950) 901.
	
	\bibitem{Vacaru}
	S. Vacaru,  Critical Remarks on Finsler Modifications of Gravity and Cosmology by Zhe Chang and Xin Li, \textit{Phys. Lett. B} \textbf{690} (2010) 224-228.
	
	\bibitem{Vacaru1}
	S. Vacaru,   Principles of Einstein–Finsler gravity and perspectives in modern cosmology,  \textit{Int. J. Mod. Phys. D} \textbf{21} (2012) 1250072.
	
	\bibitem{Schreck}
	M. Schreck,  Classical kinematics and Finsler structures for nonminimal Lorentz-violating fermions,  \textit{Eur. Phys. J. C}  \textbf{75} (2015) 187.
	
	\bibitem{Vacaru2}
	S. Vacaru, Superstrings in higher order extensions of Finsler superspaces, \textit{Nucl. Phys. B} \textbf{434} (1997) 590.
	
	\bibitem{Vacaru3}
	S. Vacaru, Spinor structures and nonlinear connections in vector bundles, generalized Lagrange and Finsler spaces,  \textit{J. Math. Phys.}  \textbf{37} (1996) 508.
	
	\bibitem{Vacaru4}
	S. Vacaru, Modified dispersion relations in Hořava–Lifshitz gravity and Finsler brane models,  \textit{Gen. Relativ. Gravit.}  \textbf{44} (2012) 1015.
	
	\bibitem{Stavrinos}
	P. Stavrinos, S. Vacaru, Cyclic and ekpyrotic universes in modified Finsler osculating gravity on tangent Lorentz bundles,  \textit{Class. Quantum Gravity}  \textbf{30} (2013) 055012.
	
	\bibitem{Rajpoot}
	S. Rajpoot, S. Vacaru, Black ring and Kerr ellipsoid—Solitonic configurations in modified Finsler gravity,  \textit{Int. J. Geom. Methods Mod. Phys.}  \textbf{12} (2015) 1550102.
	
	\bibitem{Nekouee1}
	Z. Nekouee, S.K.  Narasimhamurthy, H.M. Manjunatha, S.K. Srivastava,  Finsler–Randers model for anisotropic constant-roll inflation,  \textit{Eur. Phys. J. Plus} \textbf{137} (2022) 1388.
	
	\bibitem{Nekouee}
	Z. Nekouee, S.K.  Narasimhamurthy, Thermodynamic product formulae for Finslerian Kiselev black hole,  \textit{Eur. Phys. J. C}   \textbf{83} (2023) 723.
	
	\bibitem{Praveen}
	J. Praveen, S.K. Narasimhamurthy,  Cosmological constant roll of inflation within Finsler-Barthel-Kropina geometry: a geometric approach to early universe dynamics,  \textit{New Astron.} \textbf{108} (2024) 102187.
	
	\bibitem{Sumita}
	B. Sumita, G. Shounak, P. Nupur, R. Farook, Study of gravastars in Finslerian geometry,  \textit{Eur. Phys. J. Plus}  \textbf {135} (2020) 185.
	
	\bibitem{Vacaru5}
	S.I. Vacaru,  Finsler Black Holes Induced by Noncommutative Anholonomic Distributions in Einstein Gravity,  \textit{Class. Quantum Gravity}  \textbf{27} (2010) 105003.
	
	\bibitem{Rahman}
	F. Rahman, N. Paul, A. Banerjee, S.S. De, S. Ray, A.A. Usmani, The Finslerian wormhole models,  \textit{Eur. Phys. J. C}  \textbf{76} (2016) 246.
	
	\bibitem{Manjunatha}
	H.M. Manjunatha, S.K.  Narasimhamurthy, The wormhole model with an exponential shape function in the Finslerian framework,  \textit{Chin. J. Phys.}  \textbf{77} (2022) 1561-1578.
	
	\bibitem{Krishna}
	K.P. Dasa, D. Ujjal, Possible existence of traversable wormhole in Finsler–Randers geometry, \textit{Eur. Phys. J. C}  \textbf{83} (2023) 821.
	
	\bibitem{Bohmer}
	C.G. Bohmer, T. Harko, F.S.N. Lobo,  Conformally symmetric traversable wormholes,  \textit{Phys. Rev. D} \textbf{76} (2007) 084014.
	
	\bibitem{Bhar}
	P. Bhar, F. Rahaman, T. Manna, A. Banerjee, Wormhole supported by dark energy admitting conformal motion, \textit{Eur. Phys. J. C} \textbf{76} (2016) 708.
	
	\bibitem{Hohmann}
	M. Hohmann, C. Pfeifer, M. Raidal,  H. Veermäe,  Wormholes in conformal gravity, \textit{J. Cosmol. Astropart. Phys.}  \textbf{10} (2018) 003.
	
	\bibitem{Mustafa}
	G. Mustafa, M.F. Shamir, A. Ashraf, T.C. Xia,  Noncommutative wormholes solutions with conformal motion in the background of f(G,T) gravity, \textit{Int. J. Geom. Methods Mod. Phys.} \textbf{17} (2020) 2050103.
	
	\bibitem{Piyali}
	B. Piyali,  R. Pramit, P.K. Sahoo, Phantom energy-supported wormhole model in f(R,T) gravity assuming conformal motion, \textit{Int. J. Mod. Phys. D} \textbf{31} (2022) 2250016.
	
	
	\bibitem{Varieschi}
	G.U. Varieschi, K.L. Ault,  Wormhole geometries in fourth-order conformal Weyl gravity, \textit{Int. J. Mod. Phys. D} \textbf{25} (2016) 1650064.
	
	\bibitem{Mustafa1}
	G. Mustafa, S. Waheed, M. Zubair, T.C. Xia,  Non-commutative Wormholes Exhibiting Conformal Motion in Rastall Gravity, \textit{Chin. J. Phys.} \textbf{65} (2020) 163-176.
	
	\bibitem{Hengfei}
	Wu. Hengfei, Traversable phantom wormholes via conformal symmetry in $f(R, \phi, \chi)$ gravity, \textit{Int. J. Geom. Methods Mod. Phys.} \textbf{18} (2021) 2150210.
	
	\bibitem{Kuhfittig}
	P.K.F. Kuhfittig,  Wormholes with a barotropic equation of state admitting a one-parameter group of conformal motions, \textit{Ann. Phys.} \textbf{355} (2015) 115–120.
	
	\bibitem{Banerjee}
	A. Banerjee, K.N. Singh, M.K. Jasim, F. Rahaman, Conformally symmetric traversable wormholes in f(R,T) gravity, \textit{Ann. Phys.} \textbf{422} (2020) 168295.
	
	\bibitem{Ashraf}
	A. Ashraf, G. Mustafa, M. Ahmad, I. Hussain,  Lorentz distributed wormhole solutions in f(T) gravity with off-diagonal tetrad under conformal motions, \textit{Mod. Phys. Lett. A} \textbf{35} (2020) 2050240.
	
	\bibitem{Mustafa2}
	G. Mustafa, H. Zinnat, P.K. Sahoo, Traversable wormhole inspired by non-commutative geometries in f(Q) gravity with conformal symmetry, \textit{Ann. Phys.} \textbf{437} (2022)  168751.
	
	\bibitem{Rabia}
	S.M. Rabia, A. Israr, R. Komal,  Wormhole solutions in Rastall-like-torsion-trace gravity, \textit{Chin. J. Phys.} \textbf{82} (2023) 1-14.
	
	\bibitem{Nazavari}
	N. Nazavari, Kh. Saaidi, Energy condition of wormhole and thin shell by assuming conformal symmetry in teleparallel-Rastall gravity, \textit{Phys. Scr.} \textbf{98} (2023) 115039.
	
	\bibitem{Zubair}
	M. Zubair, S. Waheed, M. Farooq, H.A. Alkhaldi, A. Akram,  New Casimir wormholes in f(R, T) gravity admitting conformal Killing vectors, \textit{Eur. Phys. J. Plus} \textbf{138} (2023) 902.
	
	\bibitem{Waheed}
	W. Saira, M. Zubair,  A study of Yukawa–Casimir wormholes in some Rastall frameworks via conformal Killing vectors approach, \textit{Chi. J. Phys.} (2024).
	
	\bibitem{Manju}
	M. Manjunath, S.K. Narasimhamurthy, Z. Nekouee, Rajesh Kumar, Y.K. Mallikarjun,  Traversable wormholes in Finsler geometry under conformal motion,   {arXiv:2401.08498v1 [gr-qc] } 2024.
	
	\bibitem{Joharinad}
	P. Joharinad, B. Bidabad, Conformal vector fields on Finsler spaces, \textit{Differ. Geom. Appl.} \textbf{31} (2013) 33–40.
	
	\bibitem{Shen}
	Z. Shen, Lectures on Finsler Geometry, World Scientific, Singapore, 2001.
	
	\bibitem{Bao2}
	D. Bao, R.L. Bryant, S.S. Chern, Z. Shen, A Sampler of Riemann–Finsler Geometry, Cambridge University Press, New York, 2004.	
	
	\bibitem{Akbar-Zadeh}
	H. Akbar-Zadeh,  Sur les espaces de Finsler à courbures sectionnelles constantes, \textit{Acad. Roy. Belg. Bull. Cl. Sci.} \textbf{74}  (1988) 281-322.
	
	\bibitem{Li1}
	X. Li, Z. Chang, Exact solution of vacuum field equation in Finsler spacetime,  \textit{Phys. Rev. D}  \textbf{90} (2014) 06404.
	
	\bibitem{Wang}
	H.C. Wang,  On Finsler Spaces with Completely Integrable Equations of Killing,
	\textit{J. Lond. Math. Soc.}  \textbf{5} 1947 1-22.
	
	\bibitem{Rahaman2010}
	F. Rahaman, M. Jamil, R. Sharma, K. Chakraborty, A class of solutions for anisotropic stars admitting conformal motion,  \textit{Astrophys. Space Sci.} \textbf{330} (2010) 249.
	
	\bibitem{Herrera1}
	L. Herrera, J. Ponce de Leon, Isotropic and anisotropic charged spheres admitting a one‐parameter group of conformal motions, \textit{J. Math. Phys.} \textbf{26} (1985) 2303.
	
	
	\bibitem{Saiedi}
	H. Saiedi, B. Nasr Esfahani, Time-dependent wormhole solutions of f (R) theory of gravity and energy conditions,  \textit{Mod. Phys. Lett. A}  \textbf{26} (2011) 1211-1219.
	
	\bibitem{Sharif1}
	M. Sharif, S. Rani, f(T) gravity and static wormhole solutions, \textit{Mod. Phys. Lett. A}  \textbf{29} (2014) 1450137.
	
	\bibitem{Konoplya}
	R.A. Konoplya,  How to tell the shape of a wormhole by its quasinormal modes,  \textit{Phys. Lett. B}  \textbf{784} (2018) 43-49.
	
	\bibitem{Lobo}
	F.S.N. Lobo, Wormholes Warp Drives and Energy Conditions, Springer, Berlin, 2017.
	
	\bibitem{Piyali1}
	B. Piyali, R. Pramit, P.K. Sahoo, Phantom energy supported wormhole model in f(R; T) gravity assuming conformal motion, \textit{Int. J. Mod. Phys. D} \textbf{31} (2022) 2250016.
	
	\bibitem{Lynden}
	D. Lynden-Bell, J. Katz, J. Bicak, Energy and angular momentum densities of stationary gravitational fields,  \textit{Phys. Rev. D}  \textbf{75} (2007) 024040,
	
	
	\bibitem{Nandi}
	K.K. Nandi, Y.Z. Zhang, R.G. Cai, A. Panchenko, Energetics in condensate star and wormholes,  \textit{Phys. Rev. D}  \textbf{79} (2009) 024011.
	
	
	\bibitem{Raychaudri}
	A. Raychaudri, Relativistic Cosmology,  \textit{Phys. Rev.} \textbf{98} (1955) 1123.
	
	\bibitem{Nojiri}
	S. Nojiri, S.D. Odintsov, Introduction to modified gravity and gravitational alternative for dark energy, \textit{Int. J. Geom. Methods Mod. Phys.} \textbf{4} (2007) 115–145.
	
	\bibitem{Morris1988}
	M.S. Morris, K.S. Thorne, U. Yurtsever, Wormholes, Time Machines, and the Weak Energy Condition, \textit{Phys. Rev. Lett.} \textbf{61} (1988) 1446.
	
	\bibitem{Rahaman1}
	F. Rahaman, M. Kalam, K.A. Rahman, Wormhole Geometry from Real Feasible Matter Sources, \textit{Int. J. Theor. Physc.} \textbf{48} (2009) 471-475.
	
	\bibitem{Sharif}
	M. Sharif, A. Waseem, Gravitational Decoupled Anisotropic Solutions in f(G) Gravity, \textit{Eur. Phys. J. C} \textbf{78} (2018) 1326.
	
	\bibitem{Shamir}
	M.F. Shamir, Z. Asghar, A. Malik, Relativistic Krori-Barua Compact Stars in f(R, T) Gravity, \textit{Fortschr. Phys.} \textbf{70} (2022) 2200134.
	
	\bibitem{Shahzad}
	G. Mustafa, M.R. Shahzad, G. Abbas, T. Xia, Stable wormholes solutions in the background of Rastall theory,	\textit{Mod. Phys. Lett. B} \textbf{35} (2020) 2050035.
	
	
\end{thebibliography}
\end{document}